\documentstyle[eqsecnum,aps,epsf]{revtex}            
\begin{document}
\draft
\preprint{Le/9605015 -- hep-ph/9605???}
\title{Neutrino Chiral Oscillations}
\author{Stefano De Leo\thanks{e-mail: {\em deleos@le.infn.it}}
           and 
        Pietro Rotelli\thanks{e-mail: {\em rotelli@le.infn.it}} }
\address{Dipartimento di Fisica, Universit\`a degli Studi Lecce and INFN, 
Sezione di Lecce\\
via Arnesano, 73100 Lecce, Italy} 
\date{May, 1996}
\maketitle


\begin{abstract}

A localized wave packet exhibits under certain conditions zitterbewegung. A 
similar phenomenon occurs for the chirality of a massive particle. In the 
case of massive neutrinos, since they are detected via the V-A weak 
charged currents, this oscillation may even explain the ``missing'' solar 
neutrino experiments. The neutrino remains a mass eigenstate  
but contains an almost sterile right-handed component. This 
qualitative discussion opens up a number of interesting physical 
considerations. 
\end{abstract}



\section{Introduction}
\label{s1}

We are interested in chiral oscillations as a possible explanation of the 
missing solar neutrino problem because, {\em neutrinos with positive chirality 
are decoupled from the neutrino absorbing charged weak currents}. 
In recent years there has been enormous interest in the problem of the 
missing solar neutrinos~\cite{msn,msn2}. With the inclusion of 
the latest experimental results, 
approximately a third of the expected solar electron neutrino flux, within 
the energy ranges measured, is missing in earth bound 
experiments~\cite{ebe,ebe2}. 
The related search for neutrino masses was stimulated by the 
mass oscillations invoked as the most promising hypothesis
to explain the missing neutrino data. If the neutrinos 
have mass then it is possible to postulate mixing between known neutrino 
species (electron, muon and tau)~\cite{mns} or between the known neutrinos 
and higher 
mass neutrinos including sterile neutrinos~\cite{sn}. Such assumptions 
together with the 
violation of individual ($e$, $\mu$, $\tau$) leptonic number permit  
oscillations between neutrino species. This hypothesis is extensively 
studied in the literature~\cite{lit}, and is characterised by the 
parameterization 
with the differences between the squared values of neutrino masses and, 
of course, by the mixing angles. As we shall see below, our chiral  
oscillations are quite different, and do not involve any mixing angles.

Let us forget the neutrino for a moment and recall some basic facts about 
zitterbewegung~\cite{sak}. In first quantization, the relativistic 
Dirac equation 
predicts the existence for the spin $\frac{1}{2}$ particle, under certain 
conditions, of a time dependence (oscillation) in the average of the 
position variable 
(the velocity eigenvalues associated with this oscillation are $\pm c$ 
even for a massive particle). This phenomenon, known as zitterbewegung, is a 
consequence of the non commutation of $\bbox{x}$ with the Hamiltonian $H$. 
Since our calculation 
for chirality follows an almost identical line of reasoning, we shall not 
give here any details of the demonstration of zitterbewegung. We recall the 
possibility to avoid zitterbewegung by redefining the space variable via the 
Foldy-Wouthuysen transformation~\cite{fwt}. However, this introduces a very 
complicated position variable and is in contrast with the justification of 
the Darwin term in atomic physics~\cite{dt}.  In fact, the existence of 
zitterbewegung is no stranger than the existence of negative energy solutions, 
on which it depends.  We wish 
to recall that zitterbewegung is only manifest for wave functions 
with significant interference between positive and negative frequencies. 
In the following we shall talk of frequencies in plane waves and not of 
energies. Frequencies and energies coincide only for plane waves . For bound 
states and localized production the Hamiltonian is never the free particle 
one and frequency and energy no longer coincide. For example, the energy 
eigenstates of the hydrogen atom 
are not plane wave solutions and indeed contain both types of frequency, 
but, non the less, they represent definite energies. 

For an electron, model calculations~\cite{zit} predict an oscillating 
zitterbewegung frequency of around $10^{21}$ cycles per second, 
far to rapid at present for any hope 
of direct measurement. However, we mention as an aside that if the 
neutrinos are massive, one of them, probably the electron neutrino, is 
the least massive of all known fermions and as a consequence its 
zitterbewegung frequency may even be within the capacity of direct 
experimental time resolution. Even if we cannot envisage any practical
experimental test, given the neutrino's ephemeral properties.

The starting point for what follows is the observation 
that {\em the creation of a particle 
is always a localized operation}, notwithstanding the use in 
perturbation theory 
of plane waves (the asymptotic as well as unperturbed wave functions). 
For example a particle produced in the laboratory is obviously localized, 
sometimes its origin is measured to within a few microns in vertex 
detectors. Normally this fact is 
ignored or considered irrelevant. Nevertheless, the creation of a particle 
involves, even if for only short times, a non free 
Hamiltonian which fortunately 
we do not need to know for our present calculations. 
We shall simply assume the form of the wave function at creation as a 
gaussian, to ease calculations, hoping that this assumption at least 
approximates the true wave function. The average chirality of the particle 
then exhibits a time dependence similar to zitterbewegung, and it is this 
type of oscillation which we consider in this paper.


\section{Chirality time dependence}
\label{s2}

If a spin $\frac{1}{2}$ particle is massless and is produced in a chiral 
($\gamma^5$) eigenstate, 
then its chirality is a constant of motion in addition to being Lorentz 
invariant. These properties no longer hold for Dirac particles with mass. 
The free particle Hamiltonian, 
\begin{equation}
 H= -i\bbox{\alpha} \cdot \bbox{\partial} + m \beta ~ ,
\end{equation}
{\em which we shall assume  to represent the time-evolution operator for 
times subsequent to 
the creation of the spinor} ($t=0$), does not commute with the chiral 
operator $\gamma^5$ because of the $\beta$($=\gamma^0$) term in $H$. 
Specifically,
\begin{equation}
\label{dt}
-i\partial_t ~ \langle \gamma^5 \rangle  = \langle ~[H, \gamma^5]~\rangle 
= 2m ~ \langle \gamma^0 \gamma^5 \rangle
\end{equation}
where $m$ is the mass of the spinor, assumed a mass eigenstate. 
The mean value on the right hand side of the above equation is given 
explicitly by
\begin{equation}
\label{05}
 \langle \gamma^0 \gamma^5 \rangle = \int d^3 x ~\overline{\Psi}(x) 
\gamma^5 \Psi (x) ~ ,
\end{equation}
where $\Psi$ may be expanded in terms of plane wave solutions in the standard 
way:
\widetext
\begin{equation}
\Psi (x)  \equiv  \int \frac{d^3 k}{(2 \pi)^3} ~\frac{m}{E} 
~\sum_{\alpha = 1}^{2} ~[~a_{\alpha}(k)u_{\alpha}(k) e^{-ikx} + 
b_{\alpha}^{*}(k)v_{\alpha}(k) e^{ikx}~] ~ ,
\end{equation}
\noindent with $E$ which denotes the positive quantity 
\[E\equiv k^0= \sqrt{\bbox{k}^{2}+m^2}~.\] 
The normalization of the spinors,  
\begin{eqnarray*}
v^{+}_{\alpha}(k)v_{\beta}(k)=u^{+}_{\alpha}(k)u_{\beta}(k)  
&  = & \frac{E}{m} ~ \delta_{\alpha \beta}~,\\
- \overline{v}_{\alpha}(k)v_{\beta}(k)  = 
\overline{u}_{\alpha}(k) u_{\beta}(k)   
 & = & \delta_{\alpha \beta}~,
\end{eqnarray*}
and the normalization condition for $\Psi(x)$,
\[ \int d^3 x ~ \Psi^{+}(x) \Psi(x) = 1 ~, \]
impose  the following constraint on the coefficients $a_{\alpha}(k)$ and 
$b_{\alpha}(k)$  
\begin{equation}
 \int \frac{d^3 k}{(2 \pi)^3} ~\frac{m}{E} ~\sum_{\alpha} 
~[~\vert a_{\alpha}(k) \vert^2 +\vert b_{\alpha}(k) \vert^2 ~]~ = 1 ~ . 
\end{equation}

Now, as for zitterbewegung, the formula~(\ref{dt}) does not necessarily imply a 
physical effect. In it is easy to see that for wave packets made up of 
frequencies of only one given sign, the Dirac equation for the spinors 
$u(k)$ and $v(k)$ 
\begin{eqnarray*}
\left( {\not \!  k} - m \right) u(k) = 
\overline{u}(k) \left( {\not \!  k} - m \right) = 0 ~ ,\\
\left( {\not \!  k} + m \right) v(k) =
\overline{v}(k) \left( {\not \!  k} + m \right) = 0 ~ ,
\end{eqnarray*}
leads to a zero average in Eq.~(\ref{dt}). To  demonstrate this point, let us 
superpose 
only positive frequency plane waves. Computing Eq.~(\ref{05}), we find
\widetext
\begin{eqnarray}
 \langle \gamma^0 \gamma^5 \rangle  & =  & \int d^3 x ~ 
\frac{d^3 q}{(2 \pi)^3} ~
\frac{d^3 k}{(2 \pi)^3} ~
\frac{m^2}{E E'}~\sum_{\alpha , \beta} ~a^{*}_{\alpha}(q)a_{\beta}(k)
\overline{u}_{\alpha}(q) \gamma^5 u_{\beta}(k) e^{i(q-k)x}  \nonumber \\ 
 & =  & \int \frac{d^3 k}{(2 \pi)^3} ~\frac{m^2}{E^2}~\sum_{\alpha , \beta} 
~ a^{*}_{\alpha}(k)a_{\beta}(k) 
\overline{u}_{\alpha}(k) \gamma^5 u_{\beta}(k)~.  
\end{eqnarray}
\noindent where $E'\equiv q^0= \sqrt{\bbox{q}^{2}+m^2}$. Now as a consequence 
of the Gordon-like {\em chiral} identity,
\begin{equation}
\overline{u}_{\alpha}(q) \gamma^5 u_{\beta}(k)  =  
\frac{1}{2m} ~ \overline{u}_{\alpha}(q) \left( 
\not \! q \gamma^5 + \gamma^5 \! {\not \!  k} \right) u_{\beta}(k) ~,
\end{equation}
we have
\[
\overline{u}_{\alpha}(k) \gamma^5 u_{\beta}(k)  =  0 ~ ,
\]
and so we obtain a zero average for $\gamma^0 \gamma^5$. 
A similar proof 
works out for a superposition of only negative frequency plane waves.

On the other hand if both frequencies are present the interference terms 
yield an oscillation in chirality, 
\widetext
\begin{eqnarray}
 \langle \gamma^0 \gamma^5 \rangle & = &  \int d^3 x ~ 
\frac{d^3 q}{(2 \pi)^3} ~
\frac{d^3 k}{(2 \pi)^3} ~
\frac{m^2}{E E'} ~\sum_{\alpha , \beta} ~[~b_{\alpha}(q)a_{\beta}(k) 
\overline{v}_{\alpha}(q) \gamma^5 u_{\beta}(k) e^{-i(q+k)x} -~ \mbox{h.c.}~]
\nonumber \\ 
 & =  & \int \frac{d^3 k}{(2 \pi)^3} ~\frac{m^2}{E^2}~\sum_{\alpha , \beta} 
~[~ b_{\alpha}(\tilde{k})a_{\beta}(k)\overline{v}_{\alpha}(\tilde{k})
 \gamma^5 u_{\beta}(k) e^{-2iEt} -~ \mbox{h.c.} ~ ] ~ ,  
\end{eqnarray}
\noindent where $\tilde{k}=(E,- \bbox{k})$. The Gordon-like {\em chiral} 
identity for cross terms,
\begin{equation}
\overline{v}_{\alpha}(q) \gamma^5 u_{\beta}(k)  = -   
\frac{1}{2m} ~ \overline{v}_{\alpha}(q) \left( 
\not \! q \gamma^5 - \gamma^5 \! {\not \!  k} \right) u_{\beta}(k) ~,
\end{equation}
gives us, the non null equivalence
\begin{equation}
\label{cross}
\overline{v}_{\alpha}(\tilde{k}) \gamma^5 u_{\beta}(k)  = -   
\frac{E}{m} ~ \overline{v}_{\alpha}(\tilde{k}) \gamma^0 \gamma^5 u_{\beta}(k) ~.
\end{equation}
From now on we will work with the following (Pauli-Dirac) 
representation of gamma matrices
\begin{eqnarray}
\label{gr}
\gamma^0  & = & \left( \begin{array}{cc} 
       \openone_2 &  0\\
       0 &  $-$\openone_2 \end{array} \right) ~ ,  \nonumber \\  
\gamma^i  & = & \left( \begin{array}{cc} 
       0 & ~\sigma^i\\
       $-$\sigma^i & ~0 \end{array} \right) ~ ,  \\
\gamma^5 & = & \left( \begin{array}{cc} 
\      0 & \; \openone_2\\
       \openone_2 & \; 0 \end{array} \right) ~ . \nonumber  
\end{eqnarray}
In such a representation the spinor solutions are given by
\begin{eqnarray}
\label{ss}
u_{\alpha}(k) & = & \left( \frac{E+m}{2m} \right)^{1/2} ~ 
                    \left( \begin{array}{c} \chi_{\alpha} \\ \\
         \frac{\sigma^i k^i}{E+m}  ~\chi_{\alpha}
                    \end{array} \right)   ~ , \nonumber \\
\\
v_{\alpha}(k) & = & \left( \frac{E+m}{2m} \right)^{1/2} ~ 
                    \left( \begin{array}{c} 
         \frac{\sigma^i k^i}{E+m} ~\chi_{\alpha}  \\ \\ 
                    \chi_{\alpha}
                    \end{array} \right) ~ , \nonumber
\end{eqnarray}
where conventionally 
\[ \chi_{1} = \left( \begin{array}{c} 1\\ 0 \end{array} \right) ~~ , ~~
\chi_{2} = \left( \begin{array}{c} 0\\ 1 \end{array} \right) ~ . \]
Using Eqs.~(\ref{gr}) and (\ref{ss}), we easily check that
\[ -\gamma^0 \gamma^5 u_{\beta}(k) = v_{\beta}(\tilde{k}) ~ , \]
and so, combining this last result with Eq.~(\ref{cross}), we obtain
\begin{equation}
\overline{v}_{\alpha}(\tilde{k}) \gamma^5 u_{\beta}(k)  = 
- \frac{E}{m} ~ \overline{v}_{\alpha}(\tilde{k}) 
   \gamma^0 \gamma^5 u_{\beta}(k) =
\frac{E}{m} ~ \overline{v}_{\alpha}(\tilde{k}) v_{\beta}(\tilde{k}) = 
- \frac{E}{m} ~ \delta_{\alpha \beta} ~.
\end{equation}
Then Eq.~(\ref{dt}) becomes
\widetext
\begin{equation}
\label{dtm}
\partial_t \langle \gamma^5 \rangle   =  
\int \frac{d^3 k}{(2 \pi)^3} ~ \frac{m^2}{E^2}  
~\sum_{\alpha} ~[~-2iE b_{\alpha}(k) a_{\alpha}(\tilde{k}) e^{-2iEt}~ +
~\mbox{h.c.} ~] ~ .
\end{equation}
After performing the time integration, we find 
\begin{equation}
\label{5t}
\langle \gamma^5 \rangle_{(t)}   =   \langle \gamma^5 \rangle_{(0)} + 
\int \frac{d^3 k}{(2 \pi)^3} ~ \frac{m^2}{E^2}  
~\sum_{\alpha} ~[~b_{\alpha}(k) a_{\alpha}(\tilde{k}) 
\left( e^{-2iEt} -1 \right)~ +
~\mbox{h.c.} ~] ~ .
\end{equation}

Even this formula does not  necessarily imply  
a significant physical effect. For example,   
we shall demonstrate below that for a gaussian initial wave 
function  and if the initial state has average chirality zero
the oscillations of the $\pm$ chiralities cancel and there is again no 
overall oscillation. 

At this point we wish to discuss an apparent paradox. Now, for any plane wave
solution of the Dirac equation with mass, the rest-frame wave function is
always an equal mixture of both chiralities. This is easily seen by,
\begin{eqnarray}
\psi & = & \frac{1+\gamma^5}{2}~\psi + \frac{1-\gamma^5}{2}~\psi \nonumber \\
     & = & \psi_{+} + \psi_{-}~,
\end{eqnarray}
where $\psi_{+,-}$ correspond to chirality $\pm 1$ respectively. 
Furthermore, we have that 
\begin{eqnarray*}
 \not  \! k \psi_{\pm}  & = &  + m \psi_{\mp} ~, \\
 \not  \! k \psi_{\pm}  & = &  - m  \psi_{\mp} ~, 
\end{eqnarray*}
where $+m$ appears for positive energy plane waves and $-m$ for negative 
ones, and where $k$ is the 4-momentum of the particle. 
Thus, in the rest-frame ($\bbox{k}=0$) we find, since $(\gamma^0)^2=1$, that
\begin{equation} 
\vert \psi_{+} \vert^2  = \vert \psi_{-} \vert^2 ~. 
\end{equation}
Note that this result is \underline{not} Lorentz invariant since a Lorentz 
boost is not a unitary transformation. Thus, while the cross section is 
Lorentz invariant the chiral probabilities are not. This seems to suggest 
that cross-section measurements are 
chiral independent.  We seem to have an argument against the  
physical significance of chiral oscillations. 

The reply to this objection based upon the Lorentz 
invariance of the cross-section is simply that in any {\em given} 
Lorentz frame chiral oscillations are manifestly 
important because of the chiral 
projection form (V-A) of the charged weak currents. The chiral probability 
variations produced by Lorentz transformations (even if $\gamma^5$ commutes 
with the Lorentz generators) are automatically compensated by the wave function 
normalization conditions
and the Lorentz transformations of the intermediate vector bosons and 
other participating particles. 

This situation is similar to the apparent contradiction between 
the following facts. {\em In all known processes the neutrino is produced as 
an almost left-handed helicity eigenstate} because created dominantly in an 
ultra-relativistic 
state (where chirality $\sim$ helicity) by the V-A interaction and hence 
with average chirality 
close to $-1$. But, in its rest frame the (plane wave) neutrino has average 
chirality 
zero. Viewed from the neutrino's rest frame there is no preference between 
$+$ and $-$ chiralities.  Is this not 
in contradiction with the dominance of left-handed  neutrinos in nature? No, 
because while there exists frames common to almost all neutrinos in which the 
chirality is almost pure $-1$, there exist many, but non coincident frames for 
which $+1$ chirality dominates for each given neutrino. 

We conclude this Section with a discussion of another important question: 
in which frame shall we assume a given form of localization? A 
spherical gaussian assumption, suggested primarily by its simple 
integration properties, is certainly not frame 
independent. Thus if the sun is the relevant frame then we can assume, as 
we shall do below, that the neutrinos are created as almost pure negative 
chiral 
eigenstates. Chiral oscillations then occur and this in turn implies that 
a measurement of the neutrino solar flux on earth will detect an 
``apparent'' loss of neutrinos. 

Obviously, any neutrino created in the sun can reasonably be 
considered localized within the solar core at time zero. It is not even 
excluded that the relevant distance scale  for localization may be much 
smaller such as the dimensions of the nucleons within which the decaying 
quarks current gives rise to the neutrinos. In this case however we would 
have to take into account the wave function of the parent hadrons. 
In the extreme hypothesis, that the distance of localization (at creation) of 
the neutrinos is the Compton wavelength of the intermediate vector boson $W$, 
we will see that the chiral oscillation is very small.


\section{A gaussian model}
\label{s3}

Let us assume that at creation the neutrino wave function is 
given (or approximated) by a gaussian function in $\bbox{x}$ with range $d$ 
for the corresponding probability distribution, 
\begin{equation}
\label{0g}
\Psi (0, \bbox{x}) = (\pi d^{2})^{-3/4} 
\exp (- \bbox{x}^2 / 2d^2 ) ~ w ~ ,
\end{equation}
with the spinor $w$ normalized by $w^+ w=1$. Since we have already noted that 
both theoretically and experimentally 
the ultra-relativistic neutrinos are created in almost pure $-1$ chiral 
eigenstates (massive neutrinos cannot be {\em rigorously} in a chiral 
eigenstate), we assume for $w$ the {\em approximate} form
\begin{equation}
w \equiv \left( \begin{array}{c} \varphi\\ $-$\varphi \end{array} 
\right) ~ ,
\end{equation}
with $\varphi$ a two component spinor, such that $\varphi^+ \varphi=1/ 2$. 

Eq.~(\ref{0g}) may be used 
to obtain the value of the coefficients $a_{\alpha}(k)$ and 
$b_{\alpha}(k)$,
\begin{eqnarray*}
a_{\alpha}(k) & = & (4\pi d^{2})^{3/4}
\exp (- \bbox{k}^2 d^2 / 2) ~ u_{\alpha}^{+}(k) w ~ , \\
b_{\alpha}^{*}(k) & = & (4\pi d^{2})^{3/4}
\exp (- \bbox{k}^2 d^2 / 2) ~ v_{\alpha}^{+}(k) w ~ . 
\end{eqnarray*}
Note that starting with a negative chiral eigenstate we find 
$b_{\alpha}^{*}(k)=-a_{\alpha}(k)$. Eq.~(\ref{5t}) then yields the time 
dependent value of the average of $\gamma^5$,

\widetext
\begin{equation}
\langle \gamma^5 \rangle_{(t)}   =   -1 + 
\int \frac{d^3 k}{(2 \pi)^3} ~ \frac{m^2}{E^2}  
~\sum_{\alpha} ~[~a_{\alpha}^{*}(k) a_{\alpha}(\tilde{k}) 
\left( 1 - e^{-2iEt} \right)~ +
~\mbox{h.c.} ~] ~ ,
\end{equation}

\noindent which after algebraic manipulations becomes
\begin{equation}
\langle \gamma^5 \rangle_{(t)} = -1 + \mbox{NCO} ~ , 
\end{equation}
where NCO (N{\em eutrino} C{\em hiral} O{\em scillation}) indicates the 
following integral
\widetext
\begin{equation}
\label{nco}
\mbox{NCO} = 
\int \frac{d^3 k}{(2 \pi)^3} ~ \frac{m^2}{E^2} ~(4\pi d^{2})^{3/2}  
\exp (- \bbox{k}^2 d^2) ~[1 - \cos(2Et)] ~ .
\end{equation}

\noindent Alternatively we may quote the probability of finding 
the $(1-\gamma^5)/2$
component relevant for weak charged interactions, which coincides with its 
mean value:
\[
\langle - \rangle \equiv \int d^3 x ~\Psi^{+}(x) ~ \frac{1-\gamma^5}{2} ~ \Psi(x) ~.
\]
Explicitly, we have
\begin{equation}
\langle - \rangle = 1 - \frac{1}{2} ~ \mbox{NCO}~. 
\end{equation}
By changing the initial chirality to $+1$, 
\begin{equation}
w ~ \rightarrow ~ \tilde{w} \equiv \left( \begin{array}{c} \varphi\\ 
\varphi \end{array} 
\right) ~ ,
\end{equation}
we obtain $b_{\alpha}^{*}(k)=a_{\alpha}(k)$, and 
\begin{equation}
\langle + \rangle = 1 + \frac{1}{2} ~ \mbox{NCO}~. 
\end{equation}
Consequently, we find the anticipated result that for average intial chirality 
zero, there is no oscillation.

In spherical coordinates, Eq.~(\ref{nco}) becomes, after angular 
integration,
\widetext
\begin{equation}
\label{nco2}
\mbox{NCO} = \frac{4 d^3 }{\pi^{1/2}} ~
\int_{0}^{\infty} d \mbox{k} ~ \frac{(m \mbox{k})^2}{E^2} 
~ \exp (- \mbox{k}^2 d^2 ) ~[1 - \cos(2Et)] ~ ,
\end{equation}
where $\mbox{k}\equiv \vert \bbox{k} \vert$. For future considerations, 
we prefer to rewrite Eq.~(\ref{nco2}) after the 
following change of variable, $\rho = \mbox{k} d$,
\widetext
\[
\mbox{NCO} = \frac{4 }{\pi^{1/2}} ~
\int_{0}^{\infty} d \rho ~ \rho^2 
\left( \frac{\rho^2}{m^2 d^2} + 1 \right)^{-1}  \exp (-\rho^2) 
 \left\{ 1 - \cos \left[ 2mt \left( \frac{\rho^2}{m^2 d^2} + 1 \right)^{1/2} 
\right] \right\} ~.
\]

\noindent This equation is particularly simple when $md$ may be 
considered large, 
\[ \mbox{NCO} \sim 1-\cos (2mt) ~. \]
Then the expectation value for the negative chirality average becomes
\begin{equation}
\label{sim}
\langle - \rangle \sim \frac{1+\cos (2mt)}{2} \quad \quad [~ md \gg 1 ~] ~.
\end{equation}
The neutrino in this limit {\em oscillates between negative and positive 
chiral eigenvalues}. When 
in the chiral $+1$ eigenstate the neutrino only participates in elastic neutral 
current interactions and is thus in practice ``missing''. To obtain a lower 
value for the neutrino mass we must determine the value of $d$. If our 
gaussian wave function and correspondent gaussian probability for neutrino 
production is identified with the luminosity~\cite{core,core2} of the solar 
core, we find consistently a value of $d$ of $\sim 0.1~R_{\odot}$ 
(where $R_{\odot}$ is the solar radius). This yields a neutrino mass 
$\gg 3 \times 10^{-15}~\mbox{eV}$. In any model of oscillations we 
must allow for the fact  that the arrival times of the solar 
neutrinos varies. In a purely particle language (very short range 
localization) we would say that this 
depended upon the exact point of the sun in which the neutrino is produced.
If the oscillations are of a period much shorter than this uncertainty, then 
the above formula for observing the neutrino must be averaged over one or 
more cycles. The result would be the apparent loss of half of the neutrino 
flux. This is not yet incompatible with the solar neutrino data. This 
situation occurs for $md \gg 2\pi$ in agreement with the condition for 
the validity of the simplification made in Eq.~(\ref{sim}). 

Let us now consider two other ranges for $md$. For $md \ll 1$,  
we have
\widetext
\begin{equation}
\mbox{NCO} \sim  \frac{4 }{\pi^{1/2}} ~ m^2 d^2 ~
\int_{0}^{\infty} d \rho ~ \exp (-\rho^2) 
 \left[ 1 - \cos \left( 2 ~ \frac{t}{d}~ \rho \right) \right] ~.
\end{equation}

\noindent hence the conclusion that for very stringent localizations 
(e.g.~$d \sim 1/M_{W}$) the oscillating term is negligible, as anticipated 
in the previous Section.

When $ md \sim 1$ ($d \sim$ neutrino Compton wavelength), we must resort to 
a numerical calculation. However, in this case, we can restrict NCO and 
find its value for large time. To this aim, we set $md=1$ and perform 
the following change of variable 
\[ y = 2mt ~ ( \rho^2 + 1 )^{1/2} \]
in the time dependent part of NCO, explicitly 
\widetext
\[ \mbox{NCO} = {\cal I} + \frac{4 e}{\pi^{1/2}} ~
\int_{2mt}^{\infty} d y \left( \frac{1}{4 m^2 t^2} - 
\frac{1}{y^2} \right)^{1/2} \exp (-y^2/4m^2 t^2) ~ \cos y ~ ,  \]

\noindent where
\[
{\cal I} \equiv \frac{4 }{\pi^{1/2}} ~
\int_{0}^{\infty} d \rho ~ \frac{\rho^2}{\rho^2+ 1} ~ \exp (-\rho^2) ~ .
\]
With simple algebraic operations it is easy to show that NCO 
is limited by
\[{\cal I} -  \frac{1}{mt} ~ < ~ \mbox{NCO} 
~ < ~{\cal I} +  \frac{1}{mt} ~. \]
Thus, in the limit of $t\rightarrow \infty$, we obtain
\[ \lim_{t \rightarrow \infty} \mbox{NCO} ~ = ~{\cal I} 
\sim  0.48 ~ . \]
In this case the expectation value for the negative chirality average is
\[ 
\langle - \rangle \stackrel{t \rightarrow \infty}{\longrightarrow} 0.76 
 \quad \quad [~ md = 1 ~] ~ .
\]
To the extent that this limit corresponds to the peak passage of the 
neutrino flux through the earth, roughly $25 \%$ of the neutrino flux would 
be invisible.

\begin{figure}[bth]
\vspace*{.5cm}
\centerline{\epsfxsize=7.5cm \epsfysize=7.5cm \epsfbox{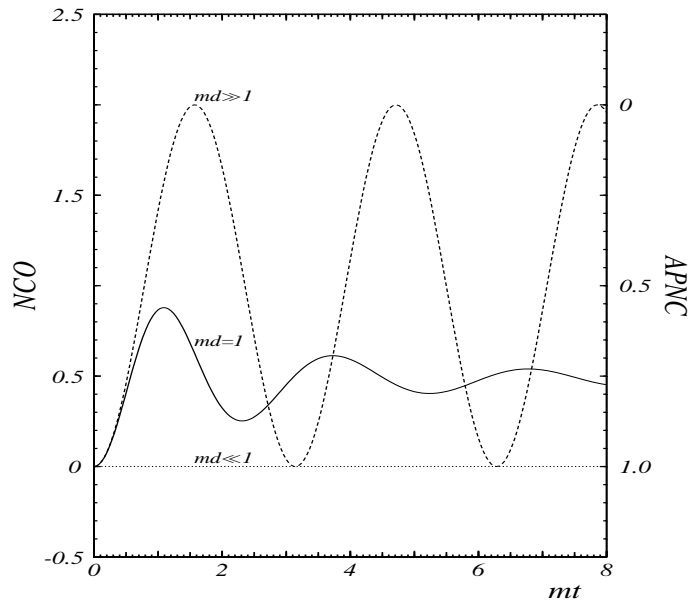}}
\vspace*{.5cm}
\caption{ The term NCO as a function of time for 
$md \gg 1$ (dashed), $md=1$ (solid) and $md \ll 1$ (dotted, practically zero). 
The right-hand scale represents $\langle - \rangle$, the Average Probability 
of Negative Chirality (APNC),  as a function of time.}
\label{fig}
\end{figure}

The results found in this Section are shown in Fig.~\ref{fig}, where NCO 
(and consequently $\langle - \rangle$) is plotted for the various cases of 
$md$ as a function of time.


\section{Conclusions}

In this work we have suggested that a spin $\frac{1}{2}$ particle produced in a 
localized condition is subject to chiral oscillations reminiscent of 
zitterbewegung. Normally this effect is of little or no interest, but for 
neutrinos whose absorption involves the V-A charged current the 
effect may be physically significant. We have suggested that this 
phenomenon may well explain the missing solar neutrino problem. Specifically 
the effect is negligible for $md \ll 1$. It predicts $\sim 25 \%$ ``loss'' for 
$md \sim 1$, and after averaging over multiple oscillations 
$\sim 50 \%$ ``loss'' for $md \gg 1$. With 
the identification of $d$ of the previous Section, the critical neutrino 
mass $m=1/d$ becomes $3 \times 10^{-15}~\mbox{eV}$. 

Only one of our basic results coincides with more traditional oscillation
models: {\em the neutrino must have a mass for this effect to exist}. 
Apart from 
this the differences are notable. Our oscillations have periods determined, 
for large $md$, by 
the particles Compton wavelength and ignores or is in addition to any 
neutrino mixing. There may be objections to the apparent non Lorentz invariance 
of our final formulas, but we have already noted that the initial localized 
form is clearly frame dependent. We recall that any result specified in a 
particular frame is by definition Lorentz invariant (and can always be 
expressed in a manifest invariant form). Explicitly, we have chosen an 
inertial Lorentz frame centered on the sun for the initial gaussian 
neutrino wave function.

We do not intend in this paper to confront 
all neutrino oscillation data. However, since the 
oscillation period is practically inversely proportional to the mass (see 
Fig.~\ref{fig}), we expect that the period of the muon neutrino be much 
smaller than that of the electron neutrino. 
Thus the atmospheric neutrino data is not incompatible with neutrino masses 
such that the atmospheric distance corresponds  
to one or more oscillations for the muon neutrino while far too short for 
any significant electron neutrino oscillation. This argument could 
be used to determine an upper limit to 
the latter mass and the order of magnitude of the muon neutrino mass, once 
the localization distance $d_{atm}$ is fixed (no longer related of course 
to the sun's core). Rather than speculate 
further on this, we wish to emphasize that the choice of an initial 
gaussian probability distribution was justified by its 
integrability, not on the basis of physical arguments and should therefore be 
treated with caution. On the other hand the considerations in this work 
invite a serious study of the appropriate creation Hamiltonian for weak 
interactions 
so as to permit the derivation of the true energy eigenvalue solutions. 
After which a quantitative calculation, even if only numerical, can be 
performed. 

We conclude with a short list of the questions which this work has stimulated. 
First, what is the creation Hamiltonian of the weak interactions? How does 
this Hamiltonian  produce a well known but often ignored fact, 
common to all created 
particles: the localization of the wave function? Furthermore, is it correct 
after the first instance to use the free Hamiltonian as the time evolution 
operator (as assumed in this work)? How are the negative frequency parts of 
the wave function to be interpreted if the reply to the 
previous question is affirmative? Finally, can all the known oscillation data 
be explained within the context of chiral oscillations?


\end{document}